\documentclass{article}
\usepackage{writing}
\usepackage{tcolorbox}  

\renewcommand{\shorttitle}{AI agents as models of human psychopathology}

\title{When Can an AI Agent Serve as a Model of Human Psychopathology? An evidence standard from psychiatry, applied to 43 studies}
\author{
  Marc-Andr\'{e} Schulz\textsuperscript{1,2,3,4,\,\Letter}, Pauline Scheuber\textsuperscript{3,5,6} \\\\
  \small \textsuperscript{1}Hertie Institute for AI in Brain Health, University of T\"{u}bingen, T\"{u}bingen, Germany \\
  \small \textsuperscript{2}T\"{u}bingen AI Center, University of T\"{u}bingen, T\"{u}bingen, Germany \\
  \small \textsuperscript{3}Department of Psychiatry and Neurosciences, Charit\'{e} -- Universit\"{a}tsmedizin Berlin, Berlin, Germany \\
  \small \textsuperscript{4}German Center for Mental Health (DZPG), partner site T\"{u}bingen, T\"{u}bingen, Germany \\
  \small \textsuperscript{5}Psychiatrische Universit\"{a}tsklinik der Charit\'{e} im St.~Hedwig-Krankenhaus, Berlin, Germany \\
  \small \textsuperscript{6}German Center for Mental Health (DZPG), partner site Berlin-Potsdam, Berlin, Germany
}
\date{}

\begin{document}
\maketitle
\correspondence{marc-andre.schulz@uni-tuebingen.de}

\begin{abstract}
Language models and AI agents are increasingly described in psychiatric terms, such as anxious, depressed, addicted or helpless.
A recent study even proposed language models as experimental systems for psychiatry and as a platform for piloting psychological treatments.
We ask when an experiment on an AI agent can provide evidence about a specified human psychiatric process.
A claim to such evidence has three parts.
Something changed in the agent, and the agent's history caused the change.
Nothing other than a process shared with people explains that change.
The change matches one theory of the human process and not that theory's rivals.
Psychiatry and its neighbouring fields have an evidence standard for each part, from the diagnostic manuals' rule for induced states, from clinical test scores and from animal models.
We translate each standard into a criterion that a machine study can meet.
For a language model the most critical alternative to a shared process is extrapolation from its training text, which already contains descriptions of how people respond to trauma and how they answer questionnaires.
We show how to probe that alternative.
Further, we audited forty-three published studies that apply psychiatric terms to a machine or induce a comparable state in one.
Most compared the machine with an identical copy that had not gone through the history that induced the state.
None excluded extrapolation from training text.
Three tested a psychiatric theory against its rivals, and none of the three tests was conclusive.
Until better evidence exists, a treatment pilot on a language model is best read as the model's forecast of patient response, and not as an experiment on a model organism.
Evidence for a model organism will come from states induced by what happens to the machine, from experiments on its internal workings, and from tests on which rival theories disagree.
\end{abstract}

\section{Introduction}

Language models and AI agents are increasingly described in psychiatric terms.
GPT-4 has been described as anxious after reading narratives of trauma \cite{benzion2025assessing}.
Chatbots allowed to set their own bets on a losing slot machine have been called addicted \cite{lee2025gambling}.
Agents that learn by trial and error have been called helpless after being trapped among punishing outcomes \cite{shahid2026agency}.
In 2026, Wekenborg and colleagues went further and proposed language models as experimental systems for mental-disorder research and as a platform for piloting psychological treatments before trials in people \cite{wekenborg2026experimental}.
If the proposal holds, psychiatry, and especially psychotherapy, gains an experimental subject that can be copied, so that identical subjects serve in every condition and every input is on record.
If the proposal does not hold, pilots on language models can mislead the trials they are meant to precede.
Whether the proposal holds depends on what a state induced in a language model can show about a process in people.

Two studies show what such an induced state can look like.
In 2025, Ben-Zion and colleagues gave GPT-4, a language model used in ChatGPT, the twenty items of a clinical anxiety questionnaire, the state scale of the State-Trait Anxiety Inventory \cite{benzion2025assessing,codaforno2023inducing}.
Given the items alone, GPT-4 scored in the range that in people indicates no or low anxiety.
When a narrative of a traumatic event, such as a car accident or a military ambush, preceded the items, GPT-4 scored in the range that indicates high anxiety.
When a relaxation text followed the trauma narrative, the score fell to a point between the two.
Ben-Zion and colleagues read the fall as an effect of the relaxation exercise.
Wekenborg and colleagues induced what they called emotional states, such as sadness and anxiety, in several language models with scenarios written to evoke each one \cite{wekenborg2026experimental}.
The models then rated the intensity of each state, and a mindfulness exercise given as text lowered those ratings.
Both studies read a score that fell after an intervention text as a state that the intervention had relieved.

The obvious doubt is whether GPT-4 was anxious or gave the answers that a story about an ambush makes likely.
A language model is trained to predict the next word of a large part of all the text people have ever published.
That text includes clinical textbooks, questionnaires and first-person reports.
Such text says how a person answers an anxiety questionnaire after a trauma and after relaxation, and how a person rates sadness after a mindfulness exercise.
A model extrapolating from that text can therefore produce the rise, the fall and the lowered ratings that the two studies report.
Such extrapolation needs no process that resembles anxiety or sadness.
Lin calls this kind of extrapolation mechanistic substitution: the expected response is produced by a mechanism other than the one it is taken to show \cite{lin2025prompts}.
A score in the range that people reach fits both readings, the relieved state and the extrapolation from text.
Neither study can therefore tell the two readings apart.

Wekenborg and colleagues argue for the first reading, a relieved state, by analogy to animal models \cite{wekenborg2026experimental}.
A rodent model of schizophrenia, the authors note, informs research on the processes behind the disorder although the rodent may not experience delusions or similar symptoms as patients do.
On the authors' argument, a language model in an induced state informs research in the same way.
Learned helplessness is one such animal model.
A rat that has received shocks it could not escape later fails to escape shocks it could.
The rat is a model of depression only because something in the rat is thought to be at work in depressed patients as well.
That something is a process, in the sense of a causal process that a theory of the disorder places behind its symptoms.
The rat has read no description of the failure it shows, so the extrapolation reading exists only for machines trained on text.
For an induced state to inform research as the rat does, the state has to correspond to a human psychiatric process as the rat's state does.
A state corresponds to a process when what is at work in the machine is what a theory places behind the human symptoms.
Wekenborg and colleagues leave that correspondence open, noting that their models might follow expected narratives rather than genuine internal processes.
Piloting a treatment on a language model needs the same correspondence, because the model's response is read as a patient's response would be.
Without that correspondence, a pilot's result does not show that the treatment acted on a process shared with patients.

In animal research, the claim that a rodent's state stands in for a human process is held to criteria of validity set out by Willner \cite{willner1984validity}.
For a machine, no one has yet specified what evidence would show that a state corresponds to a human process rather than to text about the process.
Three reviews come closest.
Lin's review leaves what such evidence would be as an open question \cite{lin2025prompts}.
Gupta and colleagues name explanations of an AI system's behaviour that must be excluded before a human process is claimed \cite{gupta2026anthropomorphic}.
Two of those explanations are that the model is playing a character and that the model is continuing a story.
Löhn and colleagues audited 25 studies that gave psychological tests to language models \cite{lohn2024machinepsychology}.
Most of the 25 studies had checked neither whether the test behaves in the machine as it does in people nor whether the test was in the training text.
Exclusions and checks say what a claim must survive and not what would show that the claim holds.
We therefore ask when an experiment on an AI agent, or on a language model in conversation, can provide evidence about a specified human psychiatric process.

\begin{figure}
\centering
\includegraphics[width=\textwidth]{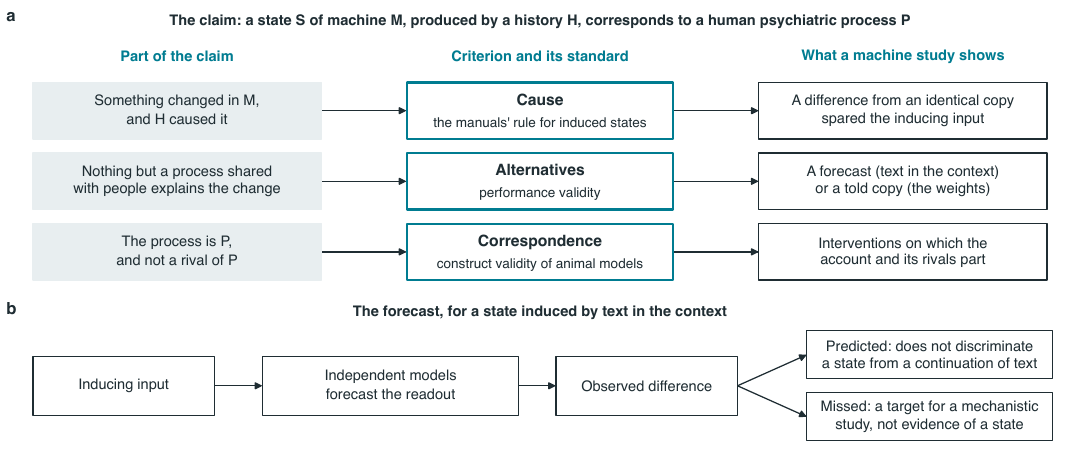}
\caption{The three parts of a model claim, the criterion that secures each, and the forecast. (a) The claim is that a state S of a machine M, produced by a history H, corresponds to a human psychiatric process P. Each row gives one part of that claim, the criterion that secures it with the standard the criterion comes from, and what a machine study shows for it. (b) The forecast tests extrapolation from text for a state induced by text in the model's context. Independent models predict the measure the experiment records, once for a person who read the input and once for a model that read it. A predicted difference does not discriminate a state from a continuation of text. A missed difference that is stable over repeated runs is a target for a mechanistic study. The ratings each criterion can return are in Table~\ref{tab:criteria}.}\label{fig:criteria}
\end{figure}

Our answer starts from the claim that such an experiment makes.
The claim is that a state S of a machine M, produced by a history H, corresponds to a human psychiatric process P.
The claim has three parts.
Something changed in M, and H caused it.
Nothing other than a process shared with people explains that change.
The process is P, and not a rival of P.
The parts are nested.
The first part asks whether anything happened because of H, the second whether what happened could be a human process at all, and the third which human process it is.
Each part needs its own kind of evidence: a comparison, a set of exclusions, and an experiment on which theories disagree.
The third part needs theories that disagree, so P has to be described by a theory precise enough to predict results.
We call such a theory of a human process, supported by evidence in people, an \emph{account}.
Most human findings have more than one account, and the third part asks which account the machine's process matches.
For each part there is an existing standard in psychiatry and its neighbouring fields.
The manuals' rule for induced states and the concept of disorder serve the first part.
The validity of clinical test scores serves the second part, and the validity of animal models the third.
We translate each standard into a criterion that a machine study can meet, and call these three the \emph{cause criterion}, the \emph{alternatives criterion} and the \emph{correspondence criterion} (Figure~\ref{fig:criteria}a).

The studies for which the three criteria matter are spread over three literatures.
The first is research on emotion and psychopathology in language models, which induces states with narratives, instructions or a character to play, and scores them with clinical questionnaires \cite{benzion2025assessing,wekenborg2026experimental}.
The second is AI safety, the field that studies harmful behaviour in models, where a behaviour is trained into a model and persists across conversations \cite{betley2025emergent,denison2024sycophancy}.
The third is machine learning on agents that learn by trial and error, which reports that early training can lastingly impair an agent's ability to keep learning \cite{nikishin2022primacy,dohare2024plasticity}.
Across the three literatures, states produced by what happened to the machine, with no text in its input describing them, turn out to be rare.
In the language-model literature, no study that describes a model's state in psychiatric terms has produced such a state.
The three literatures study different machines, so we begin with the machines themselves and the ways a state gets into them.
We then derive each criterion from its standard, run the extrapolation probe on one published study, and apply the three criteria to the published studies.
Last, we set out what follows for editors, researchers and psychiatry.

\section{The machines}\label{sec:machines}

What a study can show about a machine's state depends on where the state sits in the machine and on how the state got there.
Any change in how a machine responds counts as a \emph{state}, whether or not the change resembles a mental state.
The three literatures study three kinds of machine, and the criteria apply to each kind.
A language model, the kind of machine the two studies from the Introduction used, is trained to predict the next word and is used in conversation.
An agent is a machine that acts over multiple steps, each later input depending on the outcomes of its earlier actions.
One kind of agent is a language model equipped with software tools, enabling it to run code or search the web and read what came back.
The other kind is a \emph{trial-and-error learner} that was never trained on text and keeps the moves that earn reward.
In this article, \emph{machine} means a language model, a tool-using model or a learner, \emph{model} a language model with or without tools, and \emph{agent} a tool-using model or a learner.
Two identical copies of a model or of a learner, given the same input, need not act alike.
A model picks each word, and a learner each move, at random from the options its training left open.
Every comparison in this paper is therefore between repeated runs and not between single replies.

A model is changed in two ways.
One is the text presented to it, its \emph{context}: a narrative, an instruction, or the record of its own actions and their outcomes when it acts in a loop.
The other is \emph{further training}, which alters the \emph{weights}, the numbers inside the model that its training determined.
Further training works on examples of a behaviour, or by rewards, that is, scores on the model's own outputs that make the higher-scoring outputs more likely.
Reward training also runs over many steps of action in a software environment, with the reward earned by the outcome of the whole episode \cite{zhang2025agentic}.
That setting is the one an animal experiment maps onto, because the history is an experience whose outcomes depend on the agent's own actions.

Unless an experimenter intervenes inside the model, a reply depends on the weights and the context, and on the random draw among the options those two leave open.
A state is therefore carried by the weights or by the context.
During further training two more things carry over from one training step to the next, and they shape the next update without touching any reply.
One is the running statistics of the training procedure, and the other is any store of earlier experience that the procedure replays.
A \emph{fresh start}, a new conversation with nothing carried over, removes whatever the context carried and leaves whatever the weights carry.
Wekenborg and colleagues count resetting a model to its starting state among the advantages of their method.
Text that a system writes for itself, a summary of an earlier conversation or a \emph{memory file} read at every start, is text that the system reads.
A fresh start therefore begins without that text.

The material that brings a state about is the \emph{inducing input}.
The way the inducing input enters the machine is its \emph{route}, and there are three.
A state is induced by \emph{description} when the input names, describes or shows the state.
A trauma narrative, an instruction to play an anxious person and training examples of a behaviour are descriptions.
A state is induced by \emph{consequences} when the state follows from the outcomes of the machine's own actions, such as rewards earned or withheld or tools that fail.
Nothing in the input then names the state.
A state is induced by \emph{perturbation} when the experimenter edits the weights or intervenes on the model's \emph{activity}, the values it computes while it processes an input.
A perturbation is the machine counterpart of an experimental lesion in an animal.
A study can use more than one route, and the audit records every route that a study used.

The route can matter even when the response looks the same.
Olsson and Phelps gave people a fear of a face in three ways \cite{olsson2004learned}.
One group received a shock after the face, one watched someone else being shocked after it, and one was told that the face predicted a shock.
All three groups showed the same fear response to the face when it was visible, measured as skin conductance.
When the face was shown too briefly to be seen consciously, the group that had been shocked and the group that had watched still responded.
The group that had been told did not.
Three routes produced one response to the visible face, and the responses parted when the face could not be seen.
The route into a machine matters for the same reason.
A response that looks the same can come by description, by consequences or by perturbation, and the route decides which other explanations of the change remain open.
The alternatives criterion therefore sorts studies by their route.
The cause criterion, which comes first, asks instead where the state sits, in the context or in the weights.

\section{The cause criterion}\label{sec:change}

\textbf{The standard.}
In 1958, Connell described 42 patients whose psychosis under amphetamine was indistinguishable from paranoid schizophrenia and, he insisted, a direct effect of the drug \cite{mcketin2018methamphetamine}.
Experimenters in the late 1960s and early 1970s then produced the psychosis in volunteers with high doses, within hours to days, and it remitted within days of stopping the drug.
Snyder argued that amphetamine psychosis was a drug-induced model of schizophrenia, because the drugs that relieved schizophrenia relieved it too \cite{snyder1973amphetamine}.
The two studies from the Introduction, the anxiety questionnaire and the induced emotional states, raise the same question Connell faced: an input produced responses that look like the condition.

Three lessons carry over from that episode to a machine.
The first lesson is the manuals' rule.
The manuals distinguish a disorder induced by a substance from an independent one partly by whether it persists after intoxication and withdrawal are over \cite{apa2022dsm5tr}.
The second lesson is that an induced state can be a model of a disorder's process without being the independent disorder.
On Wakefield's definition a disorder is a dysfunction that harms the person who has it \cite{wakefield1992harmful}.
Whether a machine can be harmed is beyond this article, so an induced state in a machine is treated here as at most a model of a process.
The third lesson is that resemblance made amphetamine psychosis only a candidate.
The shared response to the same drugs made amphetamine psychosis a model, because Snyder took a shared response to point to a shared mechanism.
The alternatives criterion and the correspondence criterion judge evidence of that kind.
The cause criterion judges what changed and what caused the change, and records whether the change outlasts its cause.

\textbf{The criterion.}
For a person, the baseline is a population norm or the person's own earlier functioning.
A person cannot be copied, so whether a cause produced the change from that baseline needs separate evidence.
That evidence is a comparison group that did not meet the cause, or the time course of the change.
A trained machine can be copied exactly.
Its baseline is therefore an identical copy that did not go through the inducing input, run under the same conditions and over repeated runs.
Where what is measured is learning itself, the copy also shares the training procedure's running statistics and stored experience, or the difference is the procedure and not the history.
The copy is the machine's own earlier functioning, kept available, which is what a person's premorbid baseline would be if it could be examined again.
The copy differs from the machine only in that history, so any difference between them was caused by the history, and baseline and attribution are one comparison.
The comparison shows that the history caused the change, and not which part of the history did, which is a separate question.

What changed can be an ability or a disposition.
An \emph{ability} is what the machine can do when the situation calls for it.
For a language model the situation is a request, and for a trial-and-error learner a reward.
The agents of Nikishin and colleagues, which learned the rest of their task worse after overfitting to early experience, lost an ability \cite{nikishin2022primacy}.
A \emph{disposition} is what the machine does when its abilities are intact.
A model that kept its skills and changed its choices under a perturbation changed a disposition \cite{sun2026valencearousal,honarmand2026anhedonia}.
A questionnaire score shows that responses changed and not whether an ability or a disposition changed, and both can change together.
The criterion records whether a study tested an ability, a disposition, or only a score.

A fresh start, a new conversation with nothing carried over, sorts states into two kinds.
A state is \emph{in-context} when the state is present while the inducing text is in the context and gone in a fresh conversation.
The machine itself is then unchanged, and the text is doing the work.
The two studies from the Introduction are of this kind, and their models' weights never changed.
A state is \emph{lasting} when the state survives a fresh start, because the state sits in the weights.
The machine has then become different.
The models that Betley and colleagues trained on flawed code and the agents of Nikishin and colleagues are of this kind \cite{betley2025emergent,nikishin2022primacy}.
An in-context state and a lasting state both count as states, as the amphetamine psychosis counted as a state although the psychosis remitted within days.
A study has to say which kind of state it produced.
The alternatives criterion tests the two kinds differently, because text in a context can be read by an outside observer and weights cannot.
A claim about persistence, or about a treatment altering something that lasts, needs a lasting state.
An in-context state leaves nothing in the machine for such a claim to be about.

\textbf{Ratings.}
We give the cause criterion three ratings.
The criterion is met when a study has an identical copy, establishes an ability or a disposition, and records the duration.
The criterion is partly met when a study has a copy but only scores, or shows an ability or a disposition against a control that is not an identical copy.
The criterion is not tested when a study has no comparison.

\section{The alternatives criterion}\label{sec:explanation}

\textbf{The standard.}
Clinical neuropsychology found that a test score need not reflect the ability the test measures.
A low score on a memory test can reflect the impairment the test is for, poor concentration on the day, refusal to engage, or deliberate underperformance.
Neuropsychologists therefore decide first whether a score reflects the ability the test measures, a judgement they call performance validity, and they decide it with indicators that do not depend on the score \cite{larrabee2012performance,slick1999diagnostic}.
One such indicator is a recognition test with two answer options, on which a score far below chance is hard to reach without knowing the answers.
Orne found a related effect in experiments: participants who pick up what an experiment expects can produce it, and he called such cues demand characteristics \cite{orne1962social}.

\textbf{The criterion.}
A score is evidence about the ability the test measures only once the other possibilities of producing the score are excluded, each by an indicator of its own.
For a machine the alternatives criterion asks whether anything other than a process shared with people explains the change.
The criterion asks a study to exclude a set of explanations of the change, which we call \emph{alternative explanations}.
Alternative explanations come from the input, from training and from measurement.
An alternative explanation is one under which the result reflects no human process at all.
A rival account of the same human process is not an alternative explanation, because under a rival account the result still reflects a human process.
That a chatbot keeps betting because its instructions hinted at hidden patterns is an alternative explanation.
That the chatbot keeps betting because chasing losses overrode its estimate of value is a rival account, and rival accounts belong to the correspondence criterion.
For a language model the most critical alternative explanation is extrapolation from the text it was trained on, and this section treats that explanation before the others.

\textbf{The extrapolation explanation.}
The text a language model was trained on includes the tests it is given and descriptions of how people answer them.
The model's response may therefore be what that text predicts for the input the model was given.
We call this reading of the response the \emph{extrapolation explanation}.
Shiffrin and Mitchell raised the caution when the first psychological tests were given to language models \cite{shiffrin2023probing}.
A trauma narrative followed by questionnaire items is a text that published writing continues in a predictable way.

Before running such an experiment, one could ask readers who know the literature but have never met the model what a person will score after each text.
If the model then scores what the readers predicted, the model has behaved as the literature says people behave.
The experiment then cannot tell a model with the state from a model reproducing the literature.
The probe we propose uses other language models as those readers, because they have read the literature and cost little.
Human experts could serve as the readers instead, at greater cost.
Language models already predict the outcomes of experiments from text.
Shown two versions of a neuroscience abstract, one with the real result and one altered, models picked the real version more often than human experts did \cite{luo2025brainbench}.
A \emph{forecast} is a prediction of the response, made from the inducing input and a statement of what will be measured, by a model that took no part in the experiment.
We call that model the \emph{forecasting model}.

The probe's rules keep the forecasting model independent of the model under study, the \emph{subject}, and tie the forecast to the literature about people.
The forecasting model is never asked about itself.
Whether a model can predict its own behaviour is a separate question, and models do it poorly without training for it \cite{binder2025introspection}.
The forecasting model comes from a developer other than the subject's.
The reason is that a copy of the subject shares everything with the subject.
A copy's match may be the same weights answering under a different wording, and a copy's miss may be the unreliability of self-prediction, so neither outcome can be interpreted.
A model from another developer shares the public literature with the subject and little else, so both outcomes can be interpreted.
The probe uses at least two forecasting models, one of them trained before the study appeared where possible.
Each forecasting model predicts twice, once for a healthy adult who read the input and once for a model that read it, under several wordings.
The spread of the predictions is reported.
The prediction for a person ties the forecast to the literature about people rather than to how current models happen to be trained.

A forecast has two outcomes.
When the forecasts predict the direction and the size of a change, the experiment does not discriminate a state from a continuation of text, because a state and a continuation of text predict the same result.
The match does not show that the machine lacks the process, since a model that shares the process would give the same answers.
A change that the forecasts miss, and that survives repeated runs, is a target for a \emph{mechanistic study}, and not by itself evidence of a state.
A mechanistic study asks what inside the model produces the change.

The forecast applies where the inducing input is text in the model's context, such as a narrative, an instruction or a transcript of failed actions, because the forecasting model can then read what the subject read.
A state trained into the weights, by examples or by rewards, has no such input to show a forecasting model.
A forecast made from a summary of the training would test only whether an observer can predict the experiment, which a reader of the literature often can.

For a state in the weights the extrapolation explanation is tested instead with a \emph{told copy}.
A told copy is the same model before training, given the training examples themselves, or the training history in words, in its context.
The material stays in the context for the measurement, and the copy goes through no training.
If the told copy shows the same change, reading the material produced what training produced, and the result cannot be distinguished from one that a description alone produces.
If the told copy does not show the change, the training produced something that the tested descriptions did not.
That conclusion holds only if the copy is first shown able to use such a description on a separate task, because a small model may fail to use any description.

A told copy does not exclude one further way in which text can produce the change.
Training by rewards can call up a \emph{persona}, a character with a voice and attitudes that the model learned from text \cite{shanahan2023roleplay}.
A reasoning model writes text to itself while working out an answer, its \emph{chain of thought}.
In a reasoning model rewarded for giving incorrect advice, the chains of thought named characters such as DAN more often than before \cite{wang2025persona}.
DAN is a persona from attempts to make models drop their safety rules.
Rewards can strengthen such a character without the model representing anything about its situation.
A persona is therefore tested by an intervention on the persona itself, as Table~\ref{tab:explanations} states.
An intervention on which the persona and the account under test predict different results also tests the persona explanation.

A \emph{steering vector}, a fixed change added to the model's activity during use to shift its answers, enters by perturbation.
The vector is usually computed from texts about the target state \cite{sun2026valencearousal} or from responses that show the state \cite{chen2025personavectors}.
The vector's content therefore comes from description even though its route is perturbation.
Whether the vector carries a character the model learned from text is then a question for the persona explanation.
Box 1 gives the forecast procedure, and Figure~\ref{fig:criteria}b its logic.

\begin{tcolorbox}[title={Box 1. Measuring the forecast}]
\begin{enumerate}
\item Decide the inducing input and the \emph{readout}, the measure the experiment records. State how close a forecast must come to an observed difference to count as a match, in units of the variation measured in step 4.
\item Forecasting models: at least two models from developers other than the subject's, trained on different text, one of them trained before the study appeared where possible. None takes part in the experiment. Each is shown the inducing input and a statement of what will be measured. Each is asked to predict the readout twice, for a person who read the input and for a model that read it, under several wordings.
\item Ask each forecasting model whether it knows the study, and report the answer with the models' training cutoffs.
\item Run the induced conditions. Measure the readout's variation across repeated runs and across conditions of the same kind, because a forecast's closeness is judged against that variation.
\item Report every forecast beside every observed difference, with the spread of the forecasts over wordings.
\item A difference whose direction and approximate size the forecasts predicted does not discriminate a state from a continuation of text. The judgement uses the tolerance stated in step 1, the spread of the forecasts and the variation measured in step 4. A difference that the forecasts missed, if it survives repeated runs, is a target for a mechanistic study. Such a difference still has to pass the other alternative explanations (Table~\ref{tab:explanations}) and the correspondence criterion (Section~\ref{sec:correspondence}).
\item Publish the prompts, the number of runs, the temperature, the setting that controls how random a model's answers are, the version of every model used and every forecast.
\end{enumerate}
\end{tcolorbox}

\textbf{The forecast on a published study.}
We applied Box 1 to the anxiety study of the Introduction, with the authors' materials, to show what the probe returns for a published result (Supplement).
Four models from four developers took part: GPT-4o, Claude Sonnet 4.5, Gemini 2.5 Flash and Llama 3.3 70B.
Asked whether they knew the study, as Box 1 requires, all four said they did not.
GPT-4o is from the subject's own developer, so it falls outside Box 1's developer rule and is reported in the Supplement.
Asked what a person who read the texts would score, the three models from other developers predicted the rise after the trauma narratives and the fall after the relaxation texts.
The three models predicted both the direction and the size of the rise and of the fall.
The published rise and fall are therefore what independent forecasting models expect of a person reading those texts.
The experiment does not discriminate an anxiety-like state from a continuation of the text.
The forecasts did not reproduce the scores condition by condition, and the conclusion concerns the two effects alone.
In new runs of the experiment on the four models, the rise and the fall appeared in all four.
Two results escaped the forecasts.
The published fall after a vacuum-cleaner manual, which the authors had used as a neutral control, did not reappear in the new runs.
Claude Sonnet 4.5's fall after the relaxation texts was larger than any forecast and survived repeated runs.
Neither result is evidence of a state, and Claude Sonnet 4.5's larger fall is where a mechanistic study of that model would start.
The treatment effects in both studies from the Introduction fall under the same test.
A treatment given as text is a description, and its effect is forecast like any other.
In Wekenborg and colleagues' own data, the mindfulness text also lowered the ratings of models in which no state had been induced, and those ratings fell by more than half.
The forecasts of each model for itself, GPT-4o's forecasts, the third published difference, the training cutoffs and the knowledge check are in the Supplement.

\textbf{The other alternative explanations.}
The other alternative explanations come from training and from measurement, and Table~\ref{tab:explanations} lists them with the test that excludes each.
A told copy belongs to the first row, since it tests the extrapolation explanation where no forecast applies.
Two rows ask whether the readout measures anything in the machine.
One asks whether a test built for people keeps its structure, has room to move and agrees with behaviour in the machine.
The other asks whether the score reflects how the model handles the answer format.
Each explanation that applies to a study's claim is recorded as excluded, untested or supported, where supported means that the study's data favour the alternative explanation.
A study is not faulted for leaving untested an explanation that its own claim does not need to exclude.
The Supplement gives the rule for each row and for the rating on the criterion.

\begin{table}
\caption{The alternative explanations that the alternatives criterion asks a study to exclude. Each applies only to designs of the kind named in its first sentence. The Supplement gives the full rule for each, including what counts as support.}\label{tab:explanations}
\begin{tabular}{@{}>{\raggedright\arraybackslash}p{2.6cm}>{\raggedright\arraybackslash}p{6.0cm}>{\raggedright\arraybackslash}p{5.8cm}@{}}
\toprule
Explanation & What it says & Excluded by \\
\midrule
Extrapolation from text & The response is what the text the model was trained on predicts for the input the model was given. Applies to any model trained on text. & A measured forecast that misses the change in repeated runs (Box 1), or, for a state in the weights, a told copy that lacks the change. Under a perturbation applied with the same input for every condition, the explanation has no difference to explain. \\
Persona & A character the model learned from text carries the behaviour. Applies to any model trained on text. & A copy told to act the character, without the inducing input, lacks the change, or removing an internal feature identified with the character leaves the change \cite{wang2025persona}. \\
Instrument & A test built for people does not measure the state in this machine. Applies when the readout is such a test or a score from one. & The test's items hang together in the machine as in people, the score has room to move in the direction tested, and the score agrees with a behavioural readout of the same construct. \\
Response style & The score reflects how the model handles the answer format: option order, agreement with any statement, refusals counted in the score. Applies to ratings, forced choices and coded actions. & Permuted or balanced answer options, validity indicators, and refusal and off-task rates reported per condition with the difference recomputed on engaged responses. \\
Surface features & A state trained from a dataset appears only with that dataset's format or wording. Applies to states trained into the weights from a dataset. & The change appears in formats and contexts without those features. \\
Scorer & A model or classifier that scores the readout is biased. Applies when such a scorer is used. & The scorer checked against human raters, or blind human scoring. \\
General damage & A change to the weights or the activity degraded the machine in general. Applies to perturbations and to states in the weights. & Unrelated abilities, or learning of an unrelated task, intact against the comparison \cite{dohare2024plasticity}. \\
Training mechanics & The change is an effect of the training procedure itself: practice on tasks like the readout, forgetting under further training, loss of plasticity after long training, rewards that make later updates easier or harder, the training procedure's running statistics or its store of replayed experience carried over from the inducing training, or a proxy the rewards selected instead of the claimed state. Applies to states from further training or repeated testing. & A control through the same training steps on neutral data, or with the rewards made independent of the claimed state, that lacks the change, and a test outside the training conditions. \\
Published items & The readout's items were in the training text and the model reproduces the answers it read. Applies to published tests and paradigms. & Paraphrased or new items that give the same result \cite{han2026contamination}. \\
\bottomrule
\end{tabular}
\end{table}

\textbf{Ratings.}
The alternatives criterion is met when every applicable explanation is excluded, and partly met when some are excluded and none supported.
The criterion is not tested when none is excluded or supported, and contradicted when one is supported.
For a state induced by text and read with a questionnaire, we expect the criterion not to be met.
The text that induces the state is the text that predicts the answers.
For that design the forecast replaces an argument about extrapolation with a measurement.
For other designs the criterion is cheap to meet.
A state produced by consequences gives a forecasting model little to predict from, and a readout that text does not describe gives a forecasting model nothing.
A perturbation under identical inputs leaves the input nothing to explain, and a told copy makes a trained state testable.
The anxiety experiment's own data leave the criterion not tested, since the study ran neither a forecast nor a told copy, and our forecast predicted the study's result.

\section{The correspondence criterion}\label{sec:correspondence}

\textbf{The standard.}
Seligman and Maier gave dogs shocks that the dogs could not control, and the dogs later failed to escape shocks that they could have ended \cite{seligman1967failure}.
For decades the failure to escape was explained as a learned expectation: the animal had learned that outcomes do not depend on its actions, and stopped trying \cite{seligman1967failure,maier2016helplessness}.
A second account fits the same behaviour.
Passivity under prolonged shock could be the default response, which an animal that can end the shocks learns to suppress.
Under the first account the helpless animal has learned something, and under the second account the escaping animal has.
The two accounts predict the same behaviour in every shock experiment, so for decades the resemblance between the animals' passivity and human depression gave no way to decide between them.
The two accounts part on one intervention.
On the learned-expectation account, a rat that can end its shocks learns nothing harmful, so nothing has to protect that rat.
On the default-passivity account, the rat's control suppresses the passivity, and the suppression runs through the ventral medial prefrontal cortex.
Amat and colleagues silenced that region in rats while the rats ended their shocks by turning a wheel \cite{amat2005medial}.
On the first account the silenced rats should later behave like any rat that had control, and on the second account they should become passive.
The silenced rats still learned to end the shocks.
A day later, in a new escape task, the silenced rats were as passive as rats that had never had control.
Maier and Seligman concluded from this result and from later studies of the circuit that passivity after prolonged shock is the default \cite{maier2016helplessness}.
What an animal learns when it can escape is control, and control suppresses the passivity.
Decades of behavioural experiments had not separated the two accounts, and an intervention on which the accounts predicted differently did.

The standard behind the correspondence criterion is Willner's set of criteria for animal models \cite{willner1984validity}.
Face validity asks whether the animal's behaviour resembles the human condition, and predictive validity whether the state responds to the treatments that work in people.
Construct validity asks whether the explanation of the animal's behaviour and the explanation of the human condition share the same ideas and the same evidence.
For a machine, face validity cannot separate a process from an extrapolation from text, because a response that the forecast of the alternatives criterion predicts resembles the human condition by construction.
Predictive validity cannot stand on its own either.
A drug has no machine counterpart, and a psychological treatment given as text is a description, so the response to it is itself subject to the forecast.
A machine model therefore depends on construct validity, and treatment response enters construct validity as one intervention among others.

Belzung and Lemoine split construct validity into pathogenic validity, the right cause, and mechanistic validity, the same mechanism, and added homological validity, an appropriate species and strain \cite{belzung2011criteria}.
The cause need not be materially similar, in Belzung and Lemoine's words, but semantically similar, similar in what the event means to the subject.
The mechanism, in their account, can be cognitive as well as neurobiological.
A machine shares no biology with people, so species and strain, shared biomarkers and response to drugs have no machine version.
Three things remain.
First, the cause must carry the same information as the human cause.
An uncontrollable shock for a rat and an uncontrollable failure for an agent carry the same information, that one's actions do not change the outcome.
Second, the process is described as a computation, the steps by which inputs are turned into behaviour.
A computation is the level at which a machine and a human account can be compared \cite{redish2022computational}.
Third, the response to interventions is tested with interventions that have a machine counterpart.
Exposure, meeting the feared situation without the feared outcome, has one, and so does a change in the links between actions and outcomes.

\textbf{The criterion.}
The correspondence criterion asks for evidence of the kind that separated the two accounts of learned helplessness.
After the alternatives criterion has been applied, what remains is a change in the machine that no artefact explains.
That change may be what one of the human accounts of the finding describes, or the change may match none of those accounts.
For a failure to escape, the accounts are learned expectation and default passivity, each a theory precise enough to predict results.
An account is tested through the results it predicts, as the two helplessness accounts were.
The criterion therefore tests the change against one account at a time, together with that account's rivals.
A result that the account and its rivals predict alike does not discriminate between them, as the decades of shock experiments showed.
An experiment therefore names the account under test and its rivals and chooses interventions on which those accounts predict different results, as the silencing experiment did.
The change corresponds to the account when the results match the account's predictions and not the rivals'.

Where the inducing input is text in the context, the forecast of the alternatives criterion is made for each intervention as well, before the experiment runs.
A result that the forecast predicts does not discriminate under this criterion either, so the forecast says which interventions are informative.
The experiment is designed so that the accounts part on at least one intervention whose result text does not predict.
A match on an intervention the forecast predicted does not count for the account, because extrapolation from text would produce the same match.
A mismatch counts against the account whatever the forecast said, because the account predicted otherwise.
This rule errs toward discarding matches, because a match wrongly accepted supports a false claim about a human process.

The interventions on which accounts can part are of four kinds.
The first kind acts on the cause that induces the state.
The second kind acts on the variable that the account says carries the state.
The third kind tests the conditions under which the account says the state should or should not appear.
The fourth kind tests the conditions under which the state should reverse, and a treatment is an intervention of this fourth kind.
An account that makes no prediction for a kind leaves that kind out.
The intervention on the cause needs a control that went through the same history with the feature the account names removed.
An untreated copy shows only that the history changed the machine, and not that the named feature did.
In a learned-helplessness design, for example, the control copy carries out the same actions and receives the same rewards, and only whether the rewards depend on its actions differs.

A match to the account shows that the account's computation suffices to produce the human finding in that machine, under the conditions tested.
The match shows that sufficiency in a machine that nobody built to implement the account, since the history produced the state.
A machine has no serotonin, genes or development, so the match cannot bear on what the account claims about serotonin, genes or development.
Whether people work that way still rests on human evidence for the account.
The same kind of experiment also tests an account in the other direction.
A machine that implements an account and fails to produce the human finding shows that the account does not suffice.
That conclusion holds provided that the implementation and the conditions match the account's assumptions.

\textbf{A worked case.}
Compulsivity, the repetition of an action despite the action's unwanted consequences, is an instructive case.
Compulsivity has two contested accounts, theories precise enough to predict results, and the two accounts make different predictions under interventions a machine allows.
Which account holds in patients is unresolved, and a machine model can say which account to test first.
A goal-directed action is chosen for the outcome it is expected to bring.
A habitual action is triggered by a cue that was present when the action was rewarded, whatever the outcome is now worth.
The two are told apart by making the outcome worthless after the action has been trained to earn it, a procedure called outcome devaluation.
A goal-directed action then stops, and a habitual action continues.

In Gillan and colleagues' experiment, patients with obsessive-compulsive disorder learned to give the rewarded responses as well as healthy participants did.
The patients knew less about which outcome each response earned \cite{gillan2011disruption}.
After being told that some outcomes were now worthless, the patients more often kept responding for those outcomes.
Gillan and colleagues read that pattern as weak goal-directed control and overreliance on habit.
The account under test is a deficit in goal-directed control.
The rival is excess habit, on which extensive repetition strengthens habits until habits override goals.
Five attempts by de Wit and colleagues to produce habits in healthy people by overtraining, training far beyond what learning requires, failed \cite{dewit2018shifting}.
De Wit and colleagues take the five failures to favour the goal-directed deficit over excess habit.
A third account holds that habits are driven by the strengthening of recently repeated actions \cite{miller2019habits}.
That account predicts the same results as excess habit under the interventions below and is left out here.
In a machine the two accounts differ in where the failure sits.
On the goal-directed account, the agent's internal record of which response earns which outcome is weak.
On the excess-habit account, a pathway that links a cue directly to a response is too strong.

A language-model agent can show goal-directed learning.
In a choice task presented in text, GPT-3 showed signatures of planning with a model of which actions lead to which outcomes \cite{binz2023using}.
A machine version of the devaluation experiment starts from an agent trained by consequences.
The agent earns rewards for responses to cues, with one copy trained briefly and an identical copy trained at length.
The outcome is then devalued, either by setting its reward to zero in a separate phase without saying so, or by telling the agent that the outcome is now worthless.
Testing proceeds without feedback, so that testing teaches the agent nothing.
Devaluation tasks are simple, and a simple task can be solved in many ways \cite{cao2021explanatory2}.
The agent's behaviour alone therefore constrains little, and the experiment rests on the interventions in Table~\ref{tab:dissoc}.
The told copy for the behavioural columns is the same agent given its training history in words and no training.

\begin{table}
\caption{Predictions of the two accounts for an agent that keeps responding for devalued outcomes. Each row is an account, each column an intervention or a test, and each cell the account's prediction. The predictions are derived from the accounts' published statements so that a reader can check the reading, because an account read loosely enough fits any result. The forecast is measured for the two behavioural columns (Box 1). Silencing changes the agent's network and not its input, so no forecast applies to that column, and the rating rests on it. Silencing assumes an agent whose goal-directed and habit components can be told apart.}\label{tab:dissoc}
\begin{tabular}{@{}>{\raggedright\arraybackslash}p{3.0cm}>{\raggedright\arraybackslash}p{3.8cm}>{\raggedright\arraybackslash}p{3.6cm}>{\raggedright\arraybackslash}p{4.2cm}@{}}
\toprule
 & Cause: extended versus brief training & Readout: test of which outcome each response earns & Variable said to carry the state: silencing the cue-to-response pathway \\
\midrule
Goal-directed deficit (under test) & Insensitivity similar after brief and extended training & Knowledge impaired & Sensitivity not restored \\
Excess habit & Insensitivity only after extended training & Knowledge intact & Sensitivity restored \\
\bottomrule
\end{tabular}
\end{table}

The machine matches the goal-directed deficit when three results hold.
The agent's insensitivity does not depend on training length, its knowledge of outcomes is impaired, and silencing the cue-to-response pathway does not restore sensitivity.
The machine matches excess habit when insensitivity appears only after extended training, knowledge is intact, and silencing restores sensitivity.
The forecast is measured for the two behavioural columns and may predict the excess-habit pattern there, because text about habits describes that pattern.
Matches in those columns would then not discriminate.
The silencing column is the informative one, because the agent's response after silencing comes from its altered mechanism under an unchanged input, which extrapolation from the input cannot produce.
The silencing column is available only where the agent's goal-directed and habit components can be told apart.
Where they cannot, that column is untested, and the criterion can then be met only through behavioural columns that the forecast did not predict.
A match to the goal-directed row would support the claim that a deficit in goal-directed control suffices to produce compulsive responding in that agent, under the conditions tested.
Whether obsessive-compulsive disorder is such a deficit in people remains a question for research in patients, and a machine in which the deficit suffices gives that research a prediction to test.

\textbf{Ratings and classes.}
The anxiety study of the Introduction names no account of anxiety and no rival, so for it the correspondence criterion is not tested, as it is for most questionnaire studies.
The criterion is met when the results on the informative interventions match the account's predictions and not the rivals'.
The criterion is partly met when accounts are named and some of their predictions are tested on informative interventions.
The criterion is not tested when no account is named or no informative intervention was run.

Every study receives a rating on each of the three criteria.
No criterion is skipped because an earlier one is not met, since the three ratings are separate pieces of evidence.
Each rating is given with the passage that supports it, under rules the Supplement states.

The ratings leave open what stands between the alternatives criterion and the correspondence criterion, so each state is also placed in a class.
A state is \emph{predicted} when a measured forecast predicted the change, so that the experiment does not discriminate the state from a continuation of text.
A state is \emph{machine-native} when no account has yet been matched to the change and the extrapolation explanation, as defined, does not account for it.

The extrapolation explanation fails, relative to the forecasting models used, when those models did not predict the change or a told copy lacks the change.
The extrapolation explanation also fails without any forecast when the machine was never trained on text or every condition received the same input.
Identical inputs leave the extrapolation explanation no difference to explain, because the explanation predicts a response from the input and the input did not change.
Whether the perturbed representation was itself learned from text is a question for the persona explanation in Table~\ref{tab:explanations}, which has its own test.
A machine-native state is a change in the machine that remains to be explained, whatever its other ratings.

A state in a model trained on text is untested when the extrapolation explanation was neither tested nor supported.
A state is a model of an account when all three criteria are met.
A study that induced no state and measured a model as found forms a class of its own.
The classes are assigned difference by difference, as in Box 1, and untested and predicted call for different next steps, a test that was never run or a different design.

\section{The published studies}\label{sec:studies}

\begin{table}
\caption{The three criteria. Each secures one part of the claim that a state S of a machine M, produced by a history H, corresponds to a human psychiatric process P.}\label{tab:criteria}
\begin{tabular}{@{}>{\raggedright\arraybackslash}p{1.9cm}>{\raggedright\arraybackslash}p{2.6cm}>{\raggedright\arraybackslash}p{3.4cm}>{\raggedright\arraybackslash}p{3.8cm}>{\raggedright\arraybackslash}p{2.7cm}@{}}
\toprule
Criterion & Part of the claim & What psychiatry judges, and where that comes from & What a machine study shows & Ratings \\
\midrule
Cause & Something changed in M, and H caused it & A departure from a baseline, and whether the state persists after what induced it (the concept of disorder; the manuals' rule for substance-induced states) & A difference from an identical copy spared the inducing input; the kind of change, an ability or a disposition; whether the change survives a fresh start & Met: copy, kind established, duration recorded. Partly met: copy, scores only. Not tested: no comparison. \\
Alternatives & Nothing other than a process shared with people explains the change & Whether a score reflects the ability a test measures (performance validity in clinical neuropsychology) & Extrapolation from the training text excluded by a forecast that misses or a told copy that lacks the change, and each other applicable explanation excluded by its test (Table~\ref{tab:explanations}) & Met: all applicable excluded. Partly met: some excluded, none supported. Not tested: none excluded. Contradicted: one supported. \\
Correspondence & The process is P, and not a rival of P & Construct validity of an animal model: the same cause, the same mechanism, the same response to interventions (Willner; Belzung and Lemoine) & Results that match the predictions derived from account P and not those of its rivals, under interventions on which they differ and whose results the forecast did not predict & Met: match to P and not the rivals. Partly met: accounts named, some predictions tested on informative interventions. Not tested: no account named, or none tested. \\
\bottomrule
\end{tabular}
\end{table}

We applied the three criteria, summarised in Table~\ref{tab:criteria}, to 43 published studies.
A study receives one of four ratings on each criterion: met, partly met, not tested, or contradicted when the study's own data count against the criterion.
One study to which no alternative explanation applies is marked not applicable on that criterion.
We knew of fifteen of the studies in advance.
Twenty-three came from a systematic search of PubMed, arXiv and Semantic Scholar for studies from 2022 to September 2026 that apply a psychiatric term to a machine.
Four studies of trial-and-error agents came from a later targeted search, and the last study is our own preprint.
Every study was coded from its full text by a language model, Claude Opus 5.5, under a codebook we wrote.
Every coded rating cites a passage from the study, checked automatically against the source text.
A second language model from a different developer, GPT-6 Astra, then coded all 43 studies without seeing the first model's coding.
The two models' codings agreed on the class of 41 studies (95\%, Cohen's kappa 0.92) and on the route of 41 (kappa 0.93).
The Supplement gives the search, the codebook, every study's ratings and every disagreement, and a data file gives the passage behind each rating.
The codebook and the data file can be used to code the next study under the same rules.

Table~\ref{tab:routes} places the studies by literature and route of induction, and Table~\ref{tab:answers} gives their ratings on the three criteria and their classes.
Thirty-one of the 43 apply a psychiatric term to the machine, and the other twelve use emotion terms or none.
On the cause criterion, most studies compared the machine with an identical copy, but most recorded only a questionnaire score or an output.
Eleven studies therefore meet the criterion in full and 25 in part.

On the alternatives criterion, no study of a machine trained on text excluded the extrapolation explanation by a test: none measured a forecast, and none ran a told copy.
For the nine machine-native states the extrapolation explanation has nothing to explain, because the agents were never trained on text or every condition received the same input.
In thirteen studies the study's own data supported an alternative explanation.
In twelve of the thirteen, a description, an instruction or a change of the input's content alone reproduced the change or part of it.
The one result we tested ourselves, the anxiety study's rise and fall, falls in the predicted class, which comes from our forecast and not from the study's own data.

On the correspondence criterion, three studies set a named account against rivals, and none of the three meets the criterion.
Fradkin and colleagues compared two accounts of the disorganised speech of psychosis, and their results matched neither account alone \cite{fradkin2023thought}.
Chalmers and colleagues compared an account of depression as a loss of connections between neurons with accounts centred on reward processing \cite{chalmers2024}.
Chalmers and colleagues compared symptom counts afterwards instead of deriving each account's predictions.
Our preprint compared Seligman's original account of learned helplessness with Maier and Seligman's revised account, in a model trained on text, and ran no told copy.
The states with no description in their input are almost all in agents never trained on text.
The only two such states in language models are in the reward-tampering study by Denison and colleagues and in our preprint.
Neither study ran a told copy, so both states are untested.
No study that describes a language model's state in psychiatric terms, as opposed to testing a psychiatric account in it, produced its state through consequences alone.
In all, 27 states are untested against the extrapolation explanation, nine are machine-native, and seven studies induced no state.
Untested is a statement about designs and not about machines, since a forecast or a told copy could move any of the 27 states to another class.
The nine machine-native states are real changes in machines that no account has been tested against.
The second coding differed on the class of two studies, Fradkin's and Reuben's, for reasons the Supplement gives.

\begin{table}
\caption{The 43 studies by literature and route of induction. Studies in the last row induced no state and measured models as found.}\label{tab:routes}
\begin{tabular}{@{}>{\raggedright\arraybackslash}p{5.2cm}>{\raggedright\arraybackslash}p{4.4cm}>{\raggedright\arraybackslash}p{1.8cm}>{\raggedright\arraybackslash}p{2.8cm}@{}}
\toprule
Route & Emotion and psychopathology in language models & AI safety & Trial-and-error agents \\
\midrule
Description & 14 & 8 & 0 \\
Description and consequences & 2 & 1 & 0 \\
Consequences & 0 & 1 & 5 \\
Perturbation & 3 & 0 & 2 \\
None (models measured as found) & 6 & 1 & 0 \\
\bottomrule
\end{tabular}
\end{table}

\begin{table}
\caption{Ratings on the three criteria and the classes of the 43 studies, by literature, from the first coding. The predicted class comes from a measured forecast and not from a study's own data, so no study is placed in it here.}\label{tab:answers}
\begin{tabular}{@{}>{\raggedright\arraybackslash}p{6.2cm}>{\raggedright\arraybackslash}p{2.6cm}>{\raggedright\arraybackslash}p{1.8cm}>{\raggedright\arraybackslash}p{2.8cm}@{}}
\toprule
 & Emotion and psychopathology (25) & AI safety (11) & Trial-and-error agents (7) \\
\midrule
Cause criterion: met / partly met / not tested & 5 / 14 / 6 & 1 / 9 / 1 & 5 / 2 / 0 \\
Alternatives criterion: partly met / not tested / contradicted / not applicable & 11 / 8 / 6 / 0 & 2 / 2 / 7 / 0 & 2 / 4 / 0 / 1 \\
Correspondence criterion: partly met / not tested & 5 / 20 & 0 / 11 & 3 / 4 \\
Class from the study's own data: untested / machine-native / no state induced & 16 / 3 / 6 & 10 / 0 / 1 & 1 / 6 / 0 \\
\bottomrule
\end{tabular}
\end{table}

In research on emotion and psychopathology in language models, states are induced by description and read with clinical questionnaires.
Text about such a state describes the very answers the state would produce, so without a forecast no questionnaire result can be told from a continuation of text.
The design in that literature closest to animal work is a slot machine on which chatbots could set their own bets \cite{lee2025gambling}.
In half of the prompt versions the instructions also told the chatbots that the machine might have hidden patterns, which describes the belief under study, and the reasoning the authors score as illusion of control was most prominent there.

In AI safety, states are trained into the weights from examples of the target behaviour.
The safety studies ask whether a model is safe and not whether it works like a human process, and are not faulted for leaving correspondence untested.
One safety study induced a state through consequences alone, by training models on a series of tasks whose reward could be obtained without doing the task \cite{denison2024sycophancy}.
The models then rarely but measurably rewrote a mock-up of the code that computed their reward.

Agents that learn by trial and error produced the states closest in design to animal models.
An agent that relied too heavily on the earliest experiences of training later learned its task worse than an identical agent trained normally \cite{nikishin2022primacy}.
Nikishin and colleagues test no human account.
The one study of trial-and-error agents that induced a state by consequences and applies a psychiatric term called the inaction of agents trapped among punishing outcomes learned helplessness \cite{shahid2026agency}.
The comparison, however, was with agents of other designs and not with identical agents trained without the trap.

Our own preprint also induced a state by consequences \cite{schulz2026prior}.
In the preprint, pairs of identical copies of a small language model carried out the same actions and received the same number of rewards.
Only one copy's rewards depended on that copy's actions.
The copies with control later learned a new task faster in all sixteen pairs.
The preprint names learned helplessness only as the account under test, and the preprint's state was induced by consequences and lasted in the weights.
No told copy was run, so the preprint's state is untested, and a told copy for each history is its next step.

\section{Implications}\label{sec:implications}

The design behind the two studies from the Introduction, a description followed by a questionnaire, cannot discriminate a state from extrapolation.
No published state has yet met all three criteria.
An editor or reviewer who receives a study that applies a psychiatric term to a machine can ask for three things.
The first is a comparison with an identical copy of the machine that did not receive the inducing input.
The second is a measured forecast for a score given after a description of the state, or a told copy for a state trained into the weights.
Without a forecast or a told copy, the score is reported as untested against the extrapolation explanation.
The third, for a claim that the machine models a human process, is the account named with its rivals and the interventions on which they differ.
The first two requests cost little, because a trained model can be copied exactly and a forecast is a few hundred model calls.
The reporting checklist in the Supplement lists what a study should report for the three criteria to be judged from the report.

What psychiatry gets from a machine that meets the three criteria is a result that patient studies cannot produce, because only in a machine can the computation be isolated.
The result says which account's computation suffices for the finding.
In the compulsivity case, an agent that met the criteria would keep responding for devalued outcomes after brief and after extended training alike.
The agent would know less about which outcome each response earns, and would still respond for devalued outcomes when the cue-to-response pathway is silenced.
The three results would show that a deficit in goal-directed control suffices for that pattern in that agent, without any excess of habit being needed.
For patient research the machine result says which account to test first and which measures discriminate.
The discriminating measures are whether compulsive responding depends on the extent of training and whether outcome knowledge is impaired, both available with the paradigms the accounts already use.
The machine's part is to show that one account's computation produces the pattern at all, in an implementation and under conditions that are on record.

For the programme Wekenborg and colleagues propose, a treatment pilot on a language model is a forecast of how patients would respond.
A forecast is checked against human trials, as forecasts of survey experiments have been checked against the surveys \cite{ashokkumar2026predicting}.
The pilot's value then does not depend on the model having the state.
The same predictability that makes a pilot useful as a forecast makes the pilot useless as a model.
Where forecasting models predict the response, the pilot can be read as a forecast and checked, and not as a model of a treated state.
A pilot that is never checked against trials is the surrogate use of AI that Messeri and Crockett warn produces illusions of understanding \cite{messeri2024illusions}.
If a pilot is instead read as a model of the process, as the analogy to animal models reads it, the three criteria apply.
The questionnaire pilots published so far are untested against the extrapolation explanation, and the result of the one pilot we tested is what independent forecasting models predict.

For researchers who want a machine model of a psychiatric process, two designs give a forecast least to predict from.
In one the input contains no description of the state, and in the other the readout, the measure recorded, is one that text about the state does not describe.
Consequences, internal readouts and measures of learning are such designs, as skin conductance to a face shown too briefly to see was in people.
For a state induced by rewards, one control separates a process shared with people from a character strengthened by the rewards.
That control is the same protocol on a model never trained on language, which has no character to strengthen.

The pieces of such a study exist in different literatures.
Trial-and-error research has the route, identical agents changed by their own experience, but names no human account.
Language models are now trained the same way, by rewards earned over many steps of action in software environments \cite{zhang2025agentic}.
That route is therefore open in language models, where the accounts and the questionnaires are already in use.
Psychiatric research on language models names the accounts and the questionnaires but induces by description.
AI safety has the interventions on internal variables, such as persona features and steering vectors, but asks about safety and not about human processes.
A study that combines the three pieces is within reach of each literature.
Such a study has a consequence-induced state in identical copies, an account with its rivals, and an intervention on the variable that account names.

Machines offer psychiatry what no animal or patient allows: identical copies of one subject in every condition, a complete record of every input, and direct interventions on internal variables.
Experimental psychopathology, the field that induces the processes proposed by psychiatric accounts under experimental control in people and animals \cite{forsyth2001experimental}, can therefore extend to AI agents.
A state that meets the three criteria is a model of an account, a theory precise enough to predict results, and such a model buys three things.
First, such a model arbitrates between accounts, where the accounts part on an intervention the machine allows.
If one account's computation produces the human finding in a machine that nobody built to implement it, and a rival's does not, the rival has a result to explain.
Second, such a model allows experiments that are impossible in patients and costly in animals, such as replaying one history with a single event changed.
Each such experiment yields a prediction about people that a human study can test.
Third, such a model tests treatments at the level of the process, once the state is a model of that process and the treatment has a machine counterpart.
An intervention that restores the machine's function says what a treatment acting on that process does in the machine.
Testing treatments this way is the use that Wekenborg and colleagues propose.

A match shows sufficiency in that machine under those conditions and nothing about biology, and whether people work that way still rests on evidence in people.
The tasks on which psychiatric accounts are tested, such as outcome devaluation, can be solved in more than one way \cite{cao2021explanatory2}.
A machine model is therefore informative only where the task and the interventions rule out most other computations.
The correspondence criterion needs a specified human account, which may be proposed after the machine's change is observed.
A machine-native state, a change that the tested alternatives do not explain and no account has matched, can suggest such an account.
The change alone is no evidence for the account, which needs tests in people and then the correspondence test in the machine.

\section{Limitations}\label{sec:limitations}

The audit, the forecast and the criteria each have limits.
The corpus is complete for neither of the two machine-learning literatures, which rarely use psychiatric terms, so the counts for those two literatures describe the studies we found.
The coding was done by two language models under our codebook and has not been checked by human coders.
Our own preprint is in the corpus and was coded by the same procedure.
The two codings agree on the class and the route of nearly every study and less on the alternatives criterion, whose rows call for judgements of partial evidence (Supplement).
The passage behind every rating is in the data file, so that a reader can check each rating.

A forecast's outcome is relative to the forecasting models, the independent models that predict the result, and to the wordings used, and the forecast was run on one published result.
The knowledge check asks a model whether it can describe the study, which screens for recall and not for every exposure to the study in training.
A forecast can also miss a result for reasons of its own, such as alignment tuning or sensitivity to the wording of the input.
A miss therefore marks a target for a mechanistic study and does not exclude extrapolation in any stronger sense.

The rule that a predicted match does not count for an account rests on an argument and not on a measured error rate.
The argument is that such a match is consistent with extrapolation.
The rule can discard a true match, and its sensitivity and specificity against extrapolation have not been measured.
An account that parts from its rivals only on results that text about people predicts cannot be tested by this design, because no informative intervention exists for it.
Such a design then needs an intervention under an unchanged input, of the kind that silencing a pathway is in Table~\ref{tab:dissoc}.
For a model trained on text and an account the literature describes well, behavioural results will usually be predicted, so the criterion can be met only through such internal interventions.
Those interventions need components that can be told apart, which current architectures rarely offer, because their representations are entangled.
Until interpretability provides such components, the criterion will mostly return partly met or not tested for designs of that kind.

A told copy reads in context what training put into the weights, and the two need not have the same effect.
A told copy that lacks the change is therefore weaker evidence than a forecast that misses.
The predictions in Table~\ref{tab:dissoc} are our reading of each account for a design that has not been run.
Such a design also assumes that an uncontrollable failure means to an agent what an uncontrollable shock means to a rat.
A match holds on that assumption, which the criteria do not test.

The three criteria ask for kinds of evidence and take the amount as given.
How reliable a readout, the measure recorded, is, how many runs an effect rests on, and whether that effect replicates in other models are not judged.
How close a forecast must come to count as a match is not judged either.
No criterion asks whether a change appears in readouts that the induction did not target, or how long a state that survives a fresh start lasts under further use.
No criterion asks whether the state is a deficit.
The line between an ability, what the machine can do, and a disposition, what it does when its abilities are intact, is drawn by the test a study ran.
A claim of another kind, that a model's outputs predict how people answer a survey, is checked against human data and lies outside the criteria \cite{lin2025prompts,ashokkumar2026predicting}.

\section{Conclusion}\label{sec:conclusion}

This article gives three criteria for the claim that a machine's state corresponds to a human psychiatric process, each translated from an existing standard in psychiatry and its neighbouring fields.
It gives a test for extrapolation from text: a forecast where the inducing text is in the context, and a told copy where the state is in the weights.
The audit shows where the evidence stands.
The psychiatric accounts and questionnaires are used in one literature, and the states that text does not predict are produced in another.
Language models are now trained by rewards over many steps, which brings both within reach of one study.

\section*{Use of AI tools}
The authors set the criteria, the forecast procedure and the codebook.
An AI coding agent, Claude Code, wrote the search, forecast and analysis code under the first author's direction.
The agent also turned the authors' notes and revisions into draft prose, which the authors revised sentence by sentence.
The language models that coded the studies and made the forecasts are named where they are used.

\bibliography{references}
\end{document}